\documentclass[runningheads]{llncs}
\usepackage[T1]{fontenc}
\usepackage{graphicx} 
\usepackage{caption}    
\usepackage{subcaption} 
\usepackage{tikz}
\usetikzlibrary{shapes.geometric, arrows.meta, positioning}
\usepackage{pgfplots}
\pgfplotsset{compat=1.18}

\usepackage{fixme}
\fxsetup{status=draft}

\usepackage{algorithm2e}
\usepackage{amsmath, amssymb} 
\usepackage{listings}
\usepackage{amsmath} 

\definecolor{cborange}{RGB}{230,159,0}
\definecolor{cbblue}{RGB}{86,180,233}
\definecolor{cbgreen}{RGB}{0,158,115}
\definecolor{cbred}{RGB}{213,94,0}
\definecolor{cbpurple}{RGB}{204,121,167}

\usepackage{epstopdf}

\usetikzlibrary{external}
\tikzsetexternalprefix{figures/}

\graphicspath{{figures/}}
\usepackage[inkscapeformat=pdf]{svg}

\usepackage{asymptote}

\usepackage[frozencache=true,cachedir=minted-cache]{minted} 

\usepackage{hyperref}
\hypersetup{
    colorlinks = false,
    citecolor = red,
}
\usepackage[nameinlink]{cleveref}

\usepackage{easy-todo}

\title{Comprehensive $1/f$ Noise Model for TRNGs \\ via Fractional Brownian Motion}
\titlerunning{Comprehensive $1/f$ Noise Model for TRNGs} 
\author{Maciej Skorski}
\institute{Hubert Curien Laboratory and Czech Technical University in Prague}

\begin{document}

\maketitle

\begin{abstract}
Security of oscillatory true random number generators remains not fully understood due to insufficient understanding of complex $1/f^\alpha$ phase noise. To bridge this gap, we introduce fractional Brownian motion as a comprehensive theoretical framework, capturing power-law spectral densities from white to flicker frequency noise.

Our key contributions provide closed-form tractable solutions: (1) a quasi-renewal property showing conditional variance grows with power-law time dependence, enabling tractable leakage analysis; (2) closed-form min-entropy expressions under Gaussian phase posteriors; and (3) asymptotically unbiased Allan variance parameter estimation. 

This framework bridges physical modelling with cryptographic requirements, providing both theoretical foundations and practical calibration for oscillator-based TRNGs.
\end{abstract}

\section{Introduction}
\subsection{Background}

State-of-the-art random number generators leverage physical phenomena such as noise in electronic circuits for cryptographic applications. Oscillatory random number generators are particularly important due to their implementability in standard CMOS technology, potential for high throughput, relatively advanced theoretical understanding~\cite{BSI2024AIS31,baudetSecurityOscillatorBasedRandom2011,lubiczEntropyComputationOscillatorbased2024}, elegant yet intricate nonlinear dynamics, and adoption in security standards~\cite{BSI2024AIS31}.

In oscillatory random number generators, random bits are obtained by thresholding the instantaneous phase against a duty cycle threshold. At time $t$, the bit is generated as
\begin{align}
b_t = w( 2\pi f_0 t  + \phi_0  + \phi_t)
\end{align}
where $f_0$ is the nominal frequency, $\phi_0$ is the initial phase offset, $\phi_t$ represents stochastic phase noise, and $w$ is a $2\pi$-periodic square wave with duty cycle $\alpha$.

However, modeling oscillator noise is challenging due to complex low-frequency behavior with $1/f^{\alpha}$ spectral characteristics. Reliable entropy estimation requires accounting for complex dependencies in the phase noise process, particularly for quantifying uncertainty under information leakage. While empirical simulation methods exist, they lack guarantees for rigorous security analysis.

This work provides closed-form mathematical tools enabling efficient security evaluation. We establish formal spectral laws, derive closed-form entropy expressions, and develop a (nearly) unbiased method of recovering noise magnitude (model parameters) from experimental data.

\subsection{Contribution}
The novel contributions of this work are:
\begin{itemize}

\item \textbf{Realistic Noise Model.} We introduce (mixture of) fractional Brownian motion as phase noise model:
\begin{equation}\label{eq:rlfbm}
\phi^H(t) = \frac{1}{\Gamma(H + 1/2)} \int_0^t (t - u)^{H-1/2} dB_u
\end{equation}
This phase model turns out to capture empirically observed power-law power spectral densities $S(\omega) \propto \omega^{-2H-1}$ ranging from frequency white noise ($H = 0.5$) to frequency flicker noise ($H = 1$), providing rigorous mathematical foundations for oscillator-based TRNGs

\item \textbf{Quantified Leakage Resilience.} We establish the precise behaviour of the uncertainty: conditioned on the phase history up to time $s$, the uncertainty of predicting future phase values $\Phi_t = \phi_0 + 2\pi f_0 t + \phi_t$ ($t > s$) depends only on the time gap $\tau = (t-s)$, so that the variance grows proportionally to $\tau^{2H}$. This result enables tractable security analysis and calibration of the sampling bandwidth (determined by the relation of $\Delta$ to noise coefficients).

\item \textbf{Closed-Form Entropy Analysis.} We derive exact min-entropy expressions under any conditional Gaussian distribution of the phase, making our results applicable to all correlated Gaussian models of phase dynamics. This quantifies the security-noise variance trade-off through tractable posterior-predictive properties. We note that this closed form improves even upon the special case of white-noise-only oscillators studied recently in~\cite{lubiczEntropyComputationOscillatorbased2024}, eliminating the effort and uncertainty of Monte Carlo sampling.

\item \textbf{Parameter Estimation Framework.} We demonstrate that Allan variance is capable of accurately retrieving noise coefficients from sufficiently long phase observations (asymptotically unbiased, and in fact consistent), connecting theoretical models to hardware measurements.

\item \textbf{Supplementary Repository}. We provide a comprehensive supplementary repository \cite{skorskiFBMTRNGNoise2025} with implementation code, hardware-generated data, and symbolic computation notebooks in Wolfram Mathematica to support reproduction and evaluation of theoretical and empirical results.
\end{itemize}

The theoretical framework interisciplinary multiple analytical tools to bridge the gap between physical modelling, security analysis, and practical implementation:

\begin{itemize}
    \item \textbf{Gaussian Process Modelling and Spectral Analysis:} The introduced Gaussian process is shown to exhibit a power-law spectral density through rigorous Fourier analysis (\Cref{cor:freq_spectrum}), specifically using the Wigner-Ville non-stationary spectral density, with auxiliary results of independent interest. This establishes the relevance of fractional Brownian motion for oscillator noise modelling.

    \item \textbf{Stochastic Calculus for Leakage Analysis:} 
The quasi-renewal property (\Cref{thm:predictive-posterior}), which shows that upon conditioning the process restarts in a related manner, relies on non-trivial techniques of stochastic calculus. Yet, the final result is simple to state and naturally bounds the information leakage between current and past phase positions, enabling tractable evaluation of security.

    \item \textbf{Closed-Form Entropy and Bias:} By using periodic Gaussian distributions as the main tool \Cref{thm:periodic-norm}, we derive general closed-form expressions for entropy \Cref{thm:posterior-bit}, directly supporting quantitative security assessments.

    \item \textbf{Parameter Estimation and Empirical Validation:} For practical deployment, we study in-detail parameter estimation methods based on Allan variance (\Cref{thm:D2covar}), linking empirically measurable quantities to model parameters (noise coefficients) through explicit formulae with novel no-bias guarantees under sufficient experimental data.
\end{itemize}

This integrated approach systematically connects the mathematical properties of colored noise with practical requirements for secure true random number generation. \Cref{fig:roadmap} shows an overview of the approach.


\begin{figure}
\resizebox{0.95\linewidth}{!}{
\begin{tikzpicture}[
    box/.style={rectangle, draw=gray, fill=blue!10, text width=3cm, text centered, minimum height=1cm, rounded corners=2pt},
    toolbox/.style={rectangle, draw=gray, fill=green!10, text width=3cm, text centered, minimum height=1cm, rounded corners=2pt},
    achievementbox/.style={rectangle, draw=gray, fill=orange!20, text width=3cm, text centered, minimum height=1cm, rounded corners=2pt},
    goalbox/.style={rectangle, draw=gray, fill=red!20, text width=3cm, text centered, minimum height=1cm, rounded corners=2pt},
    arrow/.style={-{Stealth[length=3mm]}, thick, color=gray!70}
]
\node[toolbox] (gp) at (0,8.5) {GP Model};
\node[achievementbox] (powerlaw) at (-3,6.5) {Spectral\\Power Law $\star$};
\node[toolbox] (renewal) at (3,6.5) {Quasi-renewal\\Property};
\node[achievementbox] (posterior) at (0,4.5) {Tractable\\Posterior $\star$};
\node[achievementbox] (entropy) at (0,2.5) {Closed-form\\Entropy $\star$};
\node[goalbox] (security) at (0,0.5) {Security\\Guarantees $\checkmark$};
\node[achievementbox] (params) at (6,8.5) {Param Estimation $\star$};
\node (traces) at (6,6.5) {Traces};

\draw[arrow] (gp) -- (powerlaw) node[midway,above left] {\small Fourier Analysis};
\draw[arrow] (gp) -- (renewal) node[midway,above right] {\small Stochastic Calculus};
\draw[arrow] (renewal) -- (posterior) node[midway,right] {\small Analysis};
\draw[arrow] (posterior) -- (entropy) node[midway,right] {\small Periodic Gaussians};
\draw[arrow] (entropy) -- (security) node[midway,right] {\small Validation};
\draw[arrow] (traces) -- (params) node[midway,right] {\small Allan Variance};
\draw[arrow] (params) -- (gp) node[midway,above] {\small Calibration};

\node[left=0.2cm of gp] {\Cref{thm:covariance}};
\node[left=0.2cm of powerlaw] {\Cref{cor:freq_spectrum}};
\node[left=0.2cm of renewal] {\Cref{thm:predictive-posterior}};
\node[left=0.2cm of posterior] {\Cref{thm:predictive-posterior}};
\node[left=0.2cm of entropy] {\Cref{thm:posterior-bit}};
\node[left=0.2cm of security] {\Cref{cor:bandwith}};
\node[right=0.2cm of params] {\Cref{thm:D2covar}};

\draw[arrow, dashed] (entropy) -- (security);

\end{tikzpicture}
}
\label{fig:roadmap}
\caption{From Gaussian Processes to Security Guarantees}
\end{figure}

\subsection{Related Work} 

Current approaches to oscillator noise modelling include sampling power noise via inverse Fourier transforms, available in most scientific software and recently embedded in a unified TRNG simulator~\cite{pebay-peyroulaOpenTRNGOpensourceInitiative2024}. However, these methods are insufficient for security analysis. Recent work has explored Gaussian process approximations of phase noise~\cite{beneaImpactFlickerNoise2024a}, but these rely on numerically intensive inference and cannot scale to analyse leakage resilience against long observation histories. Researchers have also studied isolated properties of flicker noise in embedded devices, including attempts to measure the magnitude from traces~\cite{beneaImpactFlickerNoise2024a}.

The present work was inspired by atomic clock physicists Barnes and Allan~\cite{barnesStatisticalModelFlicker1966}, who studied flicker noise models in high-precision clocks and examined what corresponds to the $H=1$ case of our analysis. However, their treatment was heuristic and limited to spectral properties, lacking closed-form or approximate assessment of leakage-resilience.

\section{Prelimaries}

We start by establishing some notation and auxiliary facts:

\textbf{Signal Processing:} Covariance of process $X_t$ is $K_X(t_1,t_2) = \mathbf{E}[(X_{t_1} - \mathbf{E}[X_{t_1}])(X_{t_2} - \mathbb{E}[X_{t_2}])]$. Fourier transform is $\mathcal{F}_t\{f\}(\omega) = \int_{-\infty}^{\infty} f(t) e^{-i\omega t} \, dt$. Time-varying spectral density is $S_X(t,\omega) = \mathcal{F}_{\tau}\{K_X(t-\frac{\tau}{2},t+\frac{\tau}{2})\}(\omega)$, with time-average $\bar{S}_X(t,\omega) = \frac{1}{t} \int_0^t S_X(s,\omega) \, ds$ and long-time limit $\bar{S}_X(\omega) = \lim_{t \to \infty} \bar{S}_X(t,\omega)$.

\textbf{Mathematical Operations:} Finite difference of step $h$ with respect to $t$ is $\Delta_{t,h}$ and $\delta_{t,h}$. The Riemann-Liouville fractional integral of order $\alpha > 0$ is $I^{\alpha}_a f(t) = \frac{1}{\Gamma(\alpha)} \int_a^t (t-s)^{\alpha-1} f(s)\,ds$. For positive real $r$, modular arithmetic is $x \bmod r = x - r\lfloor x/r\rfloor$.

\textbf{Special Functions:} Elliptic theta function is $\vartheta_{3}(z|q)=1+2\sum_{n=1}^{\infty}q^{n^{2}}\cos(2nz)$ following \cite[\href{https://dlmf.nist.gov/20.2.E3}{(20.2.3)}]{NIST:DLMF}.  The generalized hypergeometric function is $_pF_q(a_1,\ldots,a_p; b_1,\ldots,b_q; z) = \sum_{n=0}^{\infty} \frac{(a_1)_n \cdots (a_p)_n}{(b_1)_n \cdots (b_q)_n} \frac{z^n}{n!}$ where $(a)_n = a(a+1)\cdots(a+n-1)$ is the Pochhammer symbol. Special cases include $_2F_1$ (Gauss hypergeometric) and $_1F_2$ functions appearing in our analysis.

\textbf{Entropy and Bias:} For binary random variable $b$, bias is $\mathbf{bias}(b)= |\mathbf{P}(b=1)-1/2|$ and min-entropy is $\mathbf{H}_{\infty}(b) =-\log_2(1/2+\mathbf{bias}(b))$.

\textbf{Stochastic Integration:} For Itô stochastic integrals against Brownian motion $B_t$, the integral $\int_0^T f(u) \, dB_u$ is well-defined when $f$ is square-integrable: $\int_0^T f(u)^2 \, du < \infty$. The Itô isometry states that $\mathbb{E}\left[\left(\int_0^T f(u) \, dB_u\right)^2\right] = \int_0^T f(u)^2 \, du$, and more generally $\mathbb{E}\left[\int_0^T f(u) \, dB_u \cdot \int_0^T g(u) \, dB_u\right] = \int_0^T f(u)g(u) \, du$.

\section{Results}

\subsection{Time Domain Properties}

The covariance of fractional Brownian motion admits a closed-form representation using hypergeometric functions~\cite{lim_asymptotic_1995}. This enables both theoretical analysis and efficient numerical computation through optimized implementations in scientific libraries. We will see that the most important cases of white and flicker frequency noise lead to simplified formulas involving only elementary functions that can be easily accelerated on GPU. Unlike spectral methods requiring inverse FFTs, covariance-based simulation can potentially offer computational advantages for certain parameter ranges and sample sizes, presenting an attractive alternative to current TRNG research tools~\cite{pebay-peyroulaOpenTRNGOpensourceInitiative2024}.

\begin{lemma}[Covariance of Fractional Brownian Motion]\label{thm:covariance}
For any $0\leqslant s,t$ and $H>0$ we have
\begin{align}
\mathbf{Cov}\left(\phi^H_s,\phi^H_t\right)
 = \frac{t^{H-\frac{1}{2}}s^{H+\frac{1}{2}}\,_2F_1\left(1,\frac{1}{2}-H;H+\frac{3}{2};\frac{s}{t}\right)}{\Gamma\left(H+\frac{1}{2}\right)\Gamma\left(H+\frac{3}{2}\right)},\quad s<t.
\end{align}
In particular, the variance is given by:
\begin{align}
    \mathbf{Var}[\phi^H_t] = \frac{t^{2H}}{2H\Gamma\left(H+\frac{1}{2}\right)^2}.
\end{align}
\end{lemma}

Using \Cref{thm:covariance}, the Gaussian process can be simulated by standard techniques such as applying Cholesky decomposition to the covariance matrix evaluated at specified time points (see \Cref{lst:fbm_implementation}).

\begin{lstlisting}[language=Python, 
                 basicstyle=\footnotesize\ttfamily,
                 keywordstyle=\color{blue}\bfseries,
                 commentstyle=\color{gray}\itshape,
                 stringstyle=\color{red},
                 numberstyle=\tiny\color{gray},
                 numbers=left,
                 frame=single,
                 backgroundcolor=\color{gray!5},
                 linewidth=\linewidth,
                 caption={Python implementation of FBM covariance and simulation},
                 label={lst:fbm_implementation}]
from scipy.special import hyp2f1, gamma
import numpy as np

# Covariance
def covariance(s, t, H):
    s, t = np.minimum(s,t), np.maximum(s,t)
    hypergeom = hyp2f1(1, 1/2 - H, H + 3/2, s/t)
    return 2 * s**(H+.5) * t**(H-.5) * hypergeom / \
           (gamma(H+.5)**2 * (2*H+1))

# Simulation
N = 10
xs = np.linspace(100, 200, 100)
K = covariance(xs[:,None], xs[None,:], H=1.0)
phi = np.linalg.cholesky(K) @ np.random.normal(size=(100,N))
\end{lstlisting}

\Cref{fig:traces} shows sample paths obtained using this approach.

\begin{figure}[h!]
    \centering

    \label{fig:traces}
\end{figure}

\begin{corollary}[Scale-Ratio Coordinates Covariance Representation]
Let $r = \max\{t,s\} / \min\{t,s\}$. Then for any nonnegative $t,s$ we have
$$
\mathbf{Cov}\left(\phi^H_t,\phi^H_s\right)
 = \frac{t^H s^H r^{-1/2}\,_2F_1\left(1,\frac{1}{2}-H;H+\frac{3}{2};1/r\right)}{\Gamma\left(H+\frac{1}{2}\right)\Gamma\left(H+\frac{3}{2}\right)}.
$$
In particular, the covariance is scale-affine and the process is self-similar with parameter $H$.
\end{corollary}

An elegant representation is obtained for the correlation, which depends only on the time ratio:
\begin{corollary}[Correlation]\label{cor:correlation}
For $r = \max\{t,s\} /\min\{t,s\}$  we have
$$
\mathbf{Cor}( \phi^H_t,\phi^H_s )  = \frac{4H\,_2F_1\left(1,\frac{1}{2}-H;H+\frac{3}{2};\frac{1}{r}\right)}{(2H+1)\sqrt{r}}
$$
In particular, $\mathbf{Cor}( \phi^H_t,\phi^H)s ) \sim C_H \sqrt{s/t}$ as $t/s \to \infty$, where $C_H = \frac{4H}{2H+1}$.
\end{corollary}

\begin{figure}[h!]
    \centering
    \includesvg[width=0.99\linewidth,inkscapelatex=false]{figures/correlation.svg}
    \caption{Correlation function for various Hurst parameters}
    \label{fig:correlation}
\end{figure}

\subsection{Spectral Properties}

Utilizing the derived covariance expressions, we now examine the frequency domain properties to confirm fractional Brownian motion as a well-suited TRNG noise model.

The phase spectrum is visible in experiments shown in \Cref{fig:spectrum}. However, rigorous analysis requires care since neither the process nor its increments are stationary.

\begin{figure}[h!]
    \centering
    \includesvg[width=0.99\linewidth,inkscapelatex=false]{figures/spectrum.svg}
    \caption{Power spectral density for various Hurst exponents.}
    \label{fig:spectrum}
\end{figure}

For non-stationary processes, we employ the time-varying Wigner-Ville spectrum. We establish the exact closed-form formula and precise asymptotic behavior in the large time-bandwidth regime $t\omega \gg 1$, addressing the gap in the previous approximation \cite{sithiSpectraRiemannLiouvilleFractional1995}. This proves that $\phi_H$ exhibits power-law spectrum $\omega^{-(2H+1)}$ (up to oscillations that are negligible on average).

\begin{theorem}[Spectrum of FBM]\label{thm:spectrum}
    For the phase process $\phi_H$ defined in \Cref{eq:rlfbm}, the instantaneous spectrum equals
    $$
    S_{\phi_H}(t,\omega) = 
    \frac{2^{2H+1}t^{2H+1}\,_1F_2\left(H+\frac{1}{2};H+1,H+\frac{3}{2};-t^2\omega^2\right)}{\Gamma(2H+2)}
    $$
    and the time-averaged spectrum satisfies
    $$
    \bar{S}_{\phi_H}(T,\omega) = \omega^{-2H-1}\left(1+O_H\left((T\omega)^{H-3/2}\right)\right),\quad T\omega\gg 1.
    $$
   The large-time limit  (as $T\to \infty$) is  $\omega^{-2H-1}$ for $0 < H < 3/2$.
\end{theorem}

Having characterized the spectrum of the phase noise, we now turn to the frequency noise, which is more commonly analyzed in clock stability literature and measured in hardware specifications ~\cite{riley_handbook_2008}.

To quantitatively connect phase and frequency noise, we develop a tool for evaluating spectra of differenced processes, accounting for non-stationarity. Starting from the covariance formula for differenced processes (\Cref{lemma:diff-cov}),  we establish the following elegant result featuring a frequency-dependent multiplier identical to the stationary case and a negligible second-difference non-correction term that is negligible under mild regularity conditions.

\begin{theorem}[Spectrum of Differenced Process]\label{thm:diff_spectrum}
    For any process $X_t$, the spectrum of its increment  $\Delta_h X = X_{t+h} - X_{t}$ equals:
    \begin{align*}
    S_{\Delta_h X}(t,\omega) = S_X(t+h,\omega) + S_X(t,\omega) - 2\cos(\omega h) S_X(t+h/2,\omega)
    \end{align*}
    and, more conveniently, in terms of the centred difference $\delta_h X = X_{t+h/2} - X_{t-h/2}$:
    $$
    S_{\delta_h X}(t,\omega) =  4\sin^2 (\omega h / 2) \cdot  S_X(t,\omega) + 
    \delta_{h/2}^2 S_X(t,\omega).
    $$
    Under time-averaging, the second difference correction term vanishes asymptotically, yielding:
    $$
    \bar{S}_{\delta_h X}(T,\omega) = 4\sin^2 (\omega h / 2) \bar{S}_X(T,\omega) + O(h/T)
    $$
    for bounded averaged spectra.
    \end{theorem}
The result allows for recovering many known properties, for instance, the spectrum of a differentiable stationary process differs by a factor of $\omega^2$. 
    
   Applying this to \Cref{thm:spectrum}, we deduce the spectral law of frequency noise, confirming the power law $\omega^{-2H+1}$ that encompasses white noise ($H=0.5$) and flicker frequency modulation ($H=1$) that are two dominant source of noise to consider in oscillatory TRNGs. This validates fractional Brownian motion as an excellent TRNG model.

Since instantaneous frequency is neither observable nor well-defined for many stochastic processes, we follow standard practice by using phase increments over small time intervals to represent measurable frequency fluctuations.
    \begin{corollary}[Fractional Frequency Spectrum]\label{cor:freq_spectrum}
    For the phase process $\phi^H$ defined in \Cref{eq:rlfbm} and fractional frequency $y_t = \frac{1}{2\pi f_0} \frac{\Delta \phi^H_t}{\Delta t}$ measured over time interval $\Delta t$, the time-averaged spectrum satisfies:
    $$
    \bar{S}_{y}(\omega) = \frac{4\sin^2(\omega \Delta t/2)}{(2\pi f_0 \Delta t)^2} \omega^{-2H-1}
    $$
    As $\Delta t \omega \to 0$, this approaches $(2\pi f_0)^{-2}\omega^{-2H+1}$ for $0 < H < 3/2$.
    \end{corollary}

\subsection{Leakage Resilience}

While the spectral analysis confirms that fractional Brownian motion exhibits the power-law behavior observed in real oscillators—validating our model's physical realism—spectral properties alone do not guarantee cryptographic security. Security depends critically on how the noise behaves under partial information leakage, which we analyse next.

In cryptographic applications, especially for random number generators vulnerable to side-channel attacks, it is crucial to quantify how much entropy remains after partial observation of the noise process. The key quantity is the \emph{conditional variance}:
$$
\text{Var}\left( \Phi^H_t \mid \Phi^H_{\leq s} \right), \quad s < t,
$$
which measures the unpredictability of future values given all past observations. 

It is important to note that, under the Gaussian process model, $\Phi_t$ and $\phi_t$ have identical conditional variances since they differ only by a deterministic mean function. Specifically, for any set of past observations $S$:
$$
\mathbf{Var}\left( \Phi^H_t \mid \Phi^H_{s\in S} \right) = \text{Var}\left( \phi^H_t \mid \phi^H_{s\in S} \right)
$$
Thus, from a variance perspective, an attacker knowing the full phase $\Phi_t$ is equivalent to one knowing only the noise $\phi_t$, making the analysis cleaner and simpler.

Coloured noise processes are generally non-Markovian and may have long memory, allowing an attacker to predict future values more accurately as more observations leak. Consequently, the number of leaked observations significantly impacts security. To illustrate this, consider flicker noise ($H=1$): while pairwise correlation decays as $\mathrm{Cor}(\phi^H_t, \phi^H_{t-\tau}) = 1 - O\left(\frac{\tau^2}{t^2} \log \frac{t}{\tau}\right)$ by \Cref{cor:correlation}, the variances conditioned upon all the past much quicker so that
$$\text{Var}\bigl(\phi^H_t \mid \phi^H_{\leq t-\tau}\bigr) = \Theta(\tau^2) \ll \text{Var}\bigl(\phi^H_t \mid \phi^H_{t-\tau}\bigr) = \Theta\left(\tau^2 \log \frac{t}{\tau}\right) ,$$
which shows that, as time grows, the leftover variance is only $o(1)$ fraction of what correlation method can capture.  This clearly highlights that complete observation history provides significantly more predictive power than pairwise correlations suggest—a crucial insight for TRNG security analysis.

The following analysis quantifies the leakage precisely for fractional Brownian motion models of phase noise, by establishing a closed-form formula for the conditional (co)variance.

\begin{theorem}[Renewal and Predictive Posterior]\label{thm:predictive-posterior}
Let $\phi^H_t$ be fractional Brownian motion defined in \Cref{eq:rlfbm}. Then:
\begin{enumerate}
\item For $s < t$:
$\mathbf{Var}\left( \phi^H_t \mid \phi^H_{\leq s} \right) = \mathbf{Var}(\phi^H_{t-s}) = \frac{(t-s)^{2H}}{2H\,\Gamma\left(H+\frac{1}{2}\right)^2}$

\item For $t \geq s > t_0$:
$\mathbf{Cov}\left[\phi^H_{t}, \phi^H_{s} \mid \phi^H_{\leq t_0}\right] = \mathbf{Cov}\left[\phi^H_{t-t_0}, \phi^H_{s-t_0}\right]$

\item The conditional distribution satisfies the following renewal property:
$\phi^H_t \mid \phi^H_{\leq s} \overset{d}{=} \mathbf{E}[\phi^H_t \mid \phi^H_{\leq s}] + \phi^H_{t-s}$
\end{enumerate}
\end{theorem}

\begin{corollary}[Drift Independence]
The same formulas hold with respect to $\Phi_t = \phi_0 + 2\pi t + \phi^H_t $. That is
$$
\mathbf{Cov}\left[\Phi^H_{t}, \Phi^H_{s} \mid \Phi^H_{\leq t_0}\right] = \mathbf{Cov}\left[\phi^H_{t-t_0}, \phi^H_{s-t_0}\right]
$$

\end{corollary}

\begin{remark}[Conditional Renewal]
Conditioned on past values, the process exhibits a renewal property with deterministic drift: it behaves like a fresh start of the original fractional Brownian motion plus a deterministic drift given by the conditional expectation.
\end{remark}

\subsection{Bit Bias and Entropy Rate}

Having quantified how much uncertainty remains after information leakage (\Cref{thm:predictive-posterior}), we can now determine the actual entropy available for bit generation. 

To this end, we first establish the distribution of periodic Gaussian variables, which—apart from being of independent mathematical interest—arise naturally in oscillators when accumulated phase noise causes the sampling point to wrap around clock cycle boundaries modulo $2\pi$.

\begin{theorem}[Distribution of Periodic Gaussian]\label{thm:periodic-norm}
For $Y = X \bmod r$ where $X \sim \mathsf{Norm}(\mu,\sigma^2)$
$$p_Y(y) = r^{-1} \vartheta_3\left(\pi(\mu-y)/r, e^{-2\pi^2\sigma^2/r^2}\right)$$
\end{theorem}

As phase noise variance $\sigma^2$ increases, the distribution approaches uniform, reducing bit bias (\Cref{fig:periodic-norm}).

\begin{figure}[h!]
  \centering
     \includesvg[inkscapelatex=true,width=0.95\linewidth]{figures/periodic-norm.svg}
     \caption{Periodic (wrapped) normal distribution $Y=\mathsf{Norm}(\mu,\sigma^2) \bmod 2\pi$.}
     \label{fig:periodic-norm}
\end{figure}

We now show how to quantify the bit entropy under \emph{any} Gaussian phase posterior (beyond the colored noise model considered here), completing our security analysis framework. This applies to scenarios where an attacker knows past phase observations $\Phi_{\leq s}$ and can manipulate the initial phase offset $\phi_0$.

\begin{theorem}[Bias/Entropy under Gaussian Phase Posterior]\label{thm:posterior-bit}
If the posterior distribution $\mathbf{P}( \Phi_t | \Phi_{\leq s})$ is Gaussian, then for threshold-based bit extraction at level $\alpha\pi$:
$$
\max_{\phi_0}\ \mathbf{bias}(b_t | \Phi_{\leq s},\phi_0) = \epsilon(\sigma(t|s), \alpha)
$$
where
$$\epsilon(\sigma,\alpha) = \left| \frac{1}{\pi} \int_{0}^{\alpha \pi} \vartheta_3\left( y/2, e^{-\sigma^2/2}\right) dy - \frac{1}{2} \right|$$
and $\sigma(t|s)^2 = {\mathbf{Var}[\phi_t | \phi_{\leq s}]}$ is the conditional phase noise variance (leftover uncertainty). Accordingly, the conditional-min entropy equals $$
\mathbf{H}_{\infty}(b_t | \Phi_{\leq s},\phi_0) = \log_2(1/2+\epsilon(\sigma(t|s), \alpha)).
$$

\end{theorem}

This bias directly determines the min-entropy available to legitimate users versus attackers. \Cref{fig:biases,fig:entropies} show how noise variance affects security: higher noise improves randomness quality and reduces attacker advantage.

\begin{figure}[h!]
       \includesvg[width=0.95\linewidth]{figures/biases.svg}
       \caption{Bit bias vs. phase noise variance}
       \label{fig:biases}
\end{figure}

\begin{figure}[h!]
       \includesvg[width=0.95\linewidth]{figures/entropies.svg}
       \caption{Min-entropy vs. phase noise variance}
       \label{fig:entropies}
\end{figure}

We conclude that the conditional bias or entropy can be determined with knowledge of noise coefficients. For TRNGs, the dominant sources are white noise ($H=1/2$) and flicker noise ($H=1$), with negligible contribution from other power-law components.

\begin{corollary}[Sampling Bandwidth]\label{cor:bandwith}
Suppose that phase noise follows a mixture of white and flicker frequency noise $\phi = \sum_{H\in \{1/2,1\}} c_H \phi^H_t$ with (platform-specific) coefficients $c_H$. If successive bits $b_n$ are sampled at times $t_n = n\Delta t$ with time interval $\Delta t$, then the min-entropy rate equals
$$
\mathbf{H}_{\infty}(b_{n} | b_{\leq n-1}, \phi_0)  = \epsilon(\sigma,\alpha)
$$
where 
$$\sigma^2 = \sum_{H\in \{1/2,1\}} c_H^2 \frac{(\Delta t)^{2H}}{2H\Gamma\left(H+\frac{1}{2}\right)^2}.$$
\end{corollary}

\subsection{Quasi-Stationarity and Allan Variance}

We require a technical result to evaluate covariance of differenced processes.

\begin{lemma}[Covariance of Differenced Process]\label{lemma:diff-cov}
Given a zero-mean process $X(t)$ we have for its first-difference:
$$ 
\mathbf{Cov}\left(\Delta_{t,h} X(t),\Delta_{s,h} X(s)\right)
= \Delta_{t,h} \Delta_{s,h} \mathbf{Cov}\left(X(t),X(s)\right)
$$
and for the second-difference:
$$
\mathbf{Cov}\left(\Delta^2_{t,h} X(t),\Delta^2_{s,h} X(s)\right)
= \Delta_{t,h}^2 \Delta_{s,h}^2 \mathbf{Cov}\left(X(t),X(s)\right).
$$
\end{lemma}

This result enables analysis of twice-differenced phase variance, connecting to Allan variance.

\begin{theorem}[Variance of Differenced Fractional Brownian Motion]\label{thm:D2covar}
$$
\mathbf{Var}(\Delta^2_{t,h} \phi^H_t) = h^{2H}\left(\frac{\left(4-4^H\right)\csc(H\pi)}{\Gamma(2H+1)}+O_H\left( (h/t)^{4-2H}\right)\right)
$$
In particular, for $H=1/2$ and $H=1$:
\begin{align}
\begin{aligned}
   \mathbf{Var}(\Delta^2_{t,h} \phi^1_t) & = \frac{4\log 2}{\pi}\cdot h^2 \left( 1+O_H((h/t)^2) \right) \\
   \mathbf{Var}(\Delta^2_{t,h} \phi^{1/2}_t) & = 2h\left( 1+O_H((h/t)^3) \right)
\end{aligned}
\end{align}
\end{theorem}

For sufficiently long observation times $t$, these formulae enable retrieval of noise coefficients by estimating the constants in $\phi_t \sim c_H \cdot \phi^H_t$. When multiple noise sources are present, a mixture of power functions must be fitted to the experimental data. 

To demonstrate this approach, we conducted an experiment using hardware-acquired data (1M samples from a CycloneV FPGA, collected from a 100MHz clock sampled at 104MHz), as illustrated in \Cref{fig:avar}. The figure displays the expected power-law mixture curve, where the asymptotic behaviour is directly proportional to the flicker noise coefficient.

\begin{figure}[t!]
    \centering
    \includesvg[width=0.95\linewidth]{figures/avar.svg}
    \caption{Normalized Allan Variance.}
    \label{fig:avar}
\end{figure}

\begin{figure}[t!]
\centering
\begin{tikzpicture}
\begin{axis}[
    xlabel={$H$},
    ylabel={},
    title={Fractional Brownian Motion Spectral Function},
    grid=major,
    grid style={dashed},
    axis lines=middle,
    xmin=0.0, xmax=2.01,
    ymin=0, ymax=10,
    minor y tick num=1,
    legend style={at={(1.05,1)}, anchor=north west},
    every axis plot/.append style={thick},
]

\addplot+[cbblue, thick, no markers] gnuplot [raw gnuplot, id=fbm] {
    set samples 1000;
    f(x) = ((4 - 4**x ) / sin(x*pi)) / gamma(2*x + 1);
    set xrange [0.01:1.99];
    plot f(x);
};
\addlegendentry{$\frac{(4-4^H)\csc(H\pi)}{\Gamma(2H+1)}$}

\end{axis}
\end{tikzpicture}
\caption{Constant in the Allan Variance Accummulation.}
\end{figure}

\section{Conclusion}

We established a computationally tractable framework with closed-form expressions for analysing phase noise in oscillatory TRNGs, eliminating the gaps of simulation-based approaches. Our fractional Brownian motion model provides explicit formulas for all key security metrics—variance scaling, entropy bounds, and parameter estimation. This enables direct analytical evaluation without Monte Carlo sampling or numerical integration.

In follow-up work, we explore GPU-accelerated simulation of coloured noise based on this stochastic model, leveraging the computational efficiency of our analytical approach.

\section*{Acknowledgements}

The author is grateful to Nathalie Bochard for support with hardware experiments, and to Viktor Fischer, Florent Bernard and Ángel Valle Gutiérrez for insightful discussions about power noise modelling.

\bibliographystyle{amsalpha}
\bibliography{citations}

\appendix

\section{Proofs}

\subsection{Proof of \Cref{thm:covariance}}

Since the Brownian motion has independent increments, stochastic integration specialized to deterministic functions (a special cse of Ito isometry) gives
\begin{equation}
K(t,s) = \frac{1}{\Gamma(H+1/2)^2} \int_{0}^{\min\{t,s\}} (|t-u||s-u|)^{H-1/2}\, \mathrm{d} u.
\end{equation}
Without loss of generality, because of the symmetry, suppose that $t<s$. To evaluate the integral, we substitute $z=t/s$ and $u=t \cdot v$ which yields 
\begin{equation}
K(t,s) = \frac{1}{\Gamma(H+1/2)^2} t^{H+1/2} s^{H-1/2} \int_{0}^{1} (1-v)^{H-1/2}  ( 1-z v )^{H-1/2} \, \mathrm{d} v.
\end{equation}
Taking $(a,b,c) = (1/2-H,1,H+3/2)$, so that $\frac{\Gamma(c)}{\Gamma(b)\Gamma(c-b)} = H+\frac{1}{2}$, and utilising the integral identity for hypergeometric functions 
 \cite[\href{http://dlmf.nist.gov/15.6.E1}{(15.6.1)}]{NIST:DLMF}, we arrive at the compact formula
\begin{align}
K(t,s) = \frac{H+\frac{1}{2}}{\Gamma(H+1/2)^2} \cdot t^{H+1/2} s^{H-1/2}\cdot\,  _2F_1\left( 1/2-H,1,H+3/2; \frac{t}{s} \right).
\end{align}
Finally, we note that the simpler formula for $H=1$ is the consequence of the identity
\begin{align}
\,_2F_1\left(-\frac{1}{2},1;\frac{5}{2};z\right)=\frac{3\left(\sqrt{z}(z+1)-(z-1)^2\tanh^{-1}\left(\sqrt{z}\right)\right)}{8z^{3/2}},
\end{align}
as well as $\Gamma(3/2)=\sqrt{\pi}/2$.

The variance is obtained by specializing $s=t$, and utilizing the identities $_2F_1\left(1,\frac{1}{2}-H;H+\frac{3}{2};1\right) = \frac{H+1/2}{2H}$ and $\Gamma(H+3/2)=(H+1/2)\Gamma(H+1/2)$ which follows by the general recurrence on gamma function \cite[\href{https://dlmf.nist.gov/5.5.E1}{(5.5.1)}]{NIST:DLMF}

\subsection{Proof of \Cref{thm:diff_spectrum}}
We can write
\begin{align}
S_{\Delta^h X}(t,\omega) &= \int_{-\infty}^{\infty} \left[K\left(t+h - \frac{\tau}{2}, t+h + \frac{\tau}{2}\right) + K\left(t - \frac{\tau}{2}, t + \frac{\tau}{2}\right) \right. \nonumber \\
&\left. - K\left(t+h - \frac{\tau}{2}, t + \frac{\tau}{2}\right) - K\left(t - \frac{\tau}{2}, t+h + \frac{\tau}{2}\right)\right] \mathrm{e}^{-\mathbf{i}\omega \tau } \, \mathrm{d} \tau
\end{align}

Finally, we obtain
\begin{align}
S_{\Delta^h X}(t,\omega) = S(t+h,\omega) + S(t,\omega) - S\left(t+\frac{h}{2},\omega\right)(e^{-\mathbf{i}\omega h} + e^{\mathbf{i}\omega h})
\end{align}
and so since $\omega$ is real we get
\begin{align}
S_{\Delta^h X}(t,\omega) = S(t+h,\omega) + S(t,\omega) - 2\cos (\omega h) S\left(t+\frac{h}{2},\omega\right).
\end{align}
Observe also that substituting the central differences $\delta_h X(t) = X(t+h/2) - X(t-h/2)$ yield 
$$S_{\delta_h X}(t,\omega) = \delta_{t,h/2}^2 S_X(t,\omega) + 4\sin^2(\omega h/2) S_X(t,\omega),$$
where $\delta_{t,h}^2 S_X(t) = S_X(t+h) - 2S_X(t) + S_X(t-h)$ is the second symmetric difference.

For brevity, fix $\omega$ and denote $f(t)=S_X(t,\omega)$. For time-averaged second differences, if  $F$ is the antiderivative of $f$ then:
$$\frac{1}{T}\int_0^T \delta_h^2 f(t) \, dt = \frac{\delta_h^2 F(T) - \delta_h^2 F(0)}{T}$$
Under time-averaging, the second difference correction term vanishes asymptotically. Under time-averaging, the second difference correction term vanishes asymptotically. Since $F(T) = T\bar{f}(T)$ where $\bar{f}(T) = \frac{1}{T}\int_0^T f(t)dt$, applying the discrete product rule yields  
$\delta_h^2 F(T) = T\delta_h^2 \bar{f}(T) + O(h)$ assuming that $\delta\bar{f}(T)$ is bounded.  Thus $\frac{1}{T}\int_0^T \delta_h^2 f(t) \, dt = \delta_h^2 \bar{f}(T) + O(h/T)$. 
This converges to $0$ if $\bar{f}(T)$ converge to a limit. Therefore, 
$$\bar{S}_{\delta_h X}(T,\omega) = \delta_{t,h/2}^2 \bar{S}_X(t,\omega) + 4\sin^2(\omega h/2) \bar{S}_X(t,\omega) + o(1),$$ as $T\rightarrow +\infty$.

\subsection{Proof of \Cref{thm:posterior-bit}}

    Suppose first that $\alpha=1/2$. It follows that the bit probability equals
    $\frac{1}{2\pi}\int_{I}  \vartheta_3\left(\pi (\mu-y)/r,e^{-2\pi^2\sigma^2/r^2}\right) \mbox{d} y $
    where $I$ is an interval of length $\pi$ corresponding to the "high" state of the square-wave $w$ (e.g. $w(y)=1$). It can be seen that changing the  phase  offset $\phi_0$ is equivalent to shifting the posterior mean $\mu$ of the conditional distribution by $\phi_0$ and thus to shifting the  interval $I$ by $\phi_0$. Because of the periodicity and symmetry around $0, w$ we find that maximizing over $I$ or, alternatively, over $\mu$ gives
    $$
    \epsilon = \frac{1}{2\pi}\int_{-r/4}^{r/4}  \vartheta_3\left(\pi y/r,e^{-2\pi^2\sigma^2/r^2}\right) \mbox{d} y - 1/2
    $$

Applying with $r = 2\pi$ we obtain
\begin{align}
    \epsilon = \frac{1}{2\pi} \int_{-\pi/2}^{\pi/2}  \vartheta_3\left( y/2,e^{-\sigma^2/2}\right) \mbox{d} y - 1/2
\end{align}
because of symmetry
\begin{align}
    \epsilon = \frac{1}{\pi} \int_{0}^{\pi/2}  \vartheta_3\left( y/2,e^{-\sigma^2/2}\right) \mbox{d} y - 1/2.
\end{align}
We know sketch the argument for the duty cycle $\alpha\not=\frac{1}{2}$ . We can prove similarly
\begin{align}
    \epsilon = \left| \frac{1}{\pi} \int_{0}^{\alpha \pi}  \vartheta_3\left( y/2,e^{-\sigma^2/2}\right) \mbox{d} y - 1/2 \right|.
\end{align}
To justify that the maximum is achieved for the interval centered  $0$ we consider the CDF of  the density function $G$. The bit probability equals then $G(x+a\pi)-G(x)$ depending on the moving endpoint $x$.  Since  $G' = p_Y$, extremum is at $p(x+a\pi)-p(x)=0$  which allows to conclude the claim.

\subsection{Proof of \Cref{thm:D2covar}}

\begin{proof}
We start from the variance. So
\begin{multline*}
\mathbf{Var}(\Delta^2{t,h} X_H(t) = K[h+t,h+t]+4K[t,t] + K[t-h,t-h] \\ -4K[t,t-h]-4K[t,h+t]+2K[t-h,h+t].
\end{multline*}
Then, by using \emph{Wolfram Mathematica} we find that the integral
$$
\mathbf{Var}(\Delta_{t;h}^2 X_H(t)) =O\left(h^4 t^{2H-4}\right)+h^{2H}\left(-\frac{\left(-4+4^H\right)\csc(H\pi)}{\Gamma(2H+1)}+O\left(h^4/t^4\right)\right),
$$

$$
\mathbf{Var}(\Delta_{t;h}^2 X_H(t)) =O\left(h^4 t^{2H-4}\right)+h^{2H}\left(-\frac{\left(-4+4^H\right)\csc(H\pi)}{\Gamma(2H+1)}+O\left(h^4/t^4\right)\right),
$$
with the derivation available in the supplementary repository~\cite{skorskiFBMTRNGNoise2025}.

This expression can be rewritten as 
$$
\mathbf{Var}(\Delta_{t,h}^2 X_H(t) =h^{2H}\left(-\frac{\left(-4+4^H\right)\csc(H\pi)}{\Gamma(2H+1)}+O\left( (h/t)^{4-2H}\right)\right),
$$
proving the result.

\end{proof}

\subsection{Proof of \Cref{thm:periodic-norm}}

Suppose that $r=1$, we obtain that
\begin{align}
p_Y(y) = \sum_{n\in\mathbb{Z}} p_X(y-n)
\end{align}
therefore by the functional form of the Jacobi theta function's series expansion
$$
p_Y(y)= \vartheta_3\left(\pi(\mu-y),e^{-2\pi^2\sigma^2}\right),
$$
with the derivation verified in \emph{Wolfram Mathematica} available in the supplementary repository~\cite{skorskiFBMTRNGNoise2025}.

Now, for any positive $r$ it holds that $X \bmod r = r\cdot ((X/r) \bmod 1)$ and so $Y/r = X/r \bmod 1$. 
Now we have
$$
p_{Y/r}(y) =  \vartheta_3\left(\pi(\mu/r-y),e^{-2\pi^2\sigma^2/r^2}\right),\quad 0<y<1/r
$$
by gaussianity and subsequently
$$
p_{Y} =  r^{-1}\vartheta_3\left(\pi(\mu/r-y/r),e^{-2\pi^2\sigma^2/r^2}\right),\quad 0<y<1
$$
by scaling the density $Y = r\cdot (Y/r)$.

\subsection{Proof of \Cref{thm:predictive-posterior}}

For the sake of discussing slightly more general results, denote $X_t = \phi^H_t$. 

First, we provide the lower bound on the variance (which would be already sufficient for security applications). Namely, we condition on values of $B_{\leq s}$ that represent less knowledge (as they are used to generate $X_t$):
\begin{equation}
\mathbf{Var}\left[ X_t | B_{\leq s} \right] =  \mathbf{Var}[\Gamma(H+1/2)^{-1}\int_s^t (t-u)^{H-1/2} \, dB_u]
\end{equation}

By the Itô isometry we have
\begin{equation}
\mathbf{Var}\left[ X_t | B_{\leq s} \right] =  \Gamma(H+1/2)^{-2}\int_s^t (t-u)^{2H-1} \, du
\end{equation}

Applying the substitution $u := s + w$ we find that the right-hand side equals
\begin{equation}
\Gamma(H+1/2)^{-2}\int_s^t (t-u)^{2H-1} \, du = \Gamma(H+1/2)^{-2}\int_0^{t-s} (t-s-w)^{2H-1} \, dw
\end{equation}

and again by the Itô isometry
\begin{equation}
\mathbf{Var}\left[ X_t | B_{\leq s} \right] = \mathbf{Var}[X_{t-s}].
\end{equation}

Now we prove the analogous formula for the conditional covariance. Let $t, s > t_0$. For definiteness, assume $t \geq s$. Conditioning on $B_{\leq t_0}$ we have:
\begin{equation}
\mathbf{Cov}\left[X_t, X_s \mid B_{\leq t_0}\right] = \mathbf{Cov}\left[ \Gamma(H+1/2)^{-1} \int_{t_0}^t (t-u)^{H-1/2} \, dB_u, \ \Gamma(H+1/2)^{-1} \int_{t_0}^s (s-v)^{H-1/2} \, dB_v \right]
\end{equation}

By Itô isometry, 
\begin{equation}
\mathbf{Cov}\left[X_t, X_s \mid B_{\leq t_0}\right] = \Gamma(H+1/2)^{-2} \int_{t_0}^{s} (t-u)^{H-1/2} (s-u)^{H-1/2} \, du.
\end{equation}

Now use the substitution $u = t_0 + w$, where $w \in [0, s - t_0]$, to rewrite the right-hand side as
\begin{equation}
\Gamma(H+1/2)^{-2} \int_{0}^{s - t_0} (t - t_0 - w)^{H-1/2} (s - t_0 - w)^{H-1/2} \, dw   = \mathbf{Cov}\left[X_{t-t_0},\ X_{s-t_0}\right].
\end{equation}

The far more non-trivial is to demonstrate that $B_{\leq s} $ represents same (not bigger) knowledge ($\sigma$-algebra) as $X_{\leq s} $. The solution is to demonstrate that $B_s$ can be retrieved from $X_{\leq s}$ for any $s$. 

An intuitive way of seeing this is to observe that the \Cref{eq:rlfbm} represents a fractional integral of order $H+1/2$ which can be (hopefully) differentiated back. The challenge is in the formal construction of a stochastic derivative, as $X_s$ fails to meet standard regularity asusmptions (bounded variation / seminmatigale).  A beter solution is to write down the integral tranform in the Volterra form:
$$
X_t = \int_0^{T}  K(t,s)  \mbox{d} B_s
$$
with the kernel $K(t,s) = C_H\cdot (t-s)^{H-1/2}\mathbf{1}\{t\geq s\}$ lower-triangular, and then to seek an analogous - inverting - integral representation of $B_t$ in terms of $X_s$
$$
B_s = \int_0^s L(s,t) \mbox{d} X_t.
$$
For details we refer to \cite{mishuraGaussianVolterraProcesses2022,sottinenCONDITIONALMEANHEDGINGTRANSACTION2018} but here sketch the argument. For the (well-defined) integral against $X_t$ substituting gives
$$
B_s = \int_0^s \int_0^t L(s,t) \frac{\partial}{\partial t} K(t,u) \ \mbox{d} t \ \mbox{d} B_u 
$$
for every $s$ and thus
$$
\int_u^t L(t,s) \frac{\partial}{\partial t} K(u,t) \ \mbox{d} t  = 1,
$$
which can be solved in the regime $0<H$ with beta functions. With formalities filled, it follows that
$$
\sigma( B_{\leq s} ) = \sigma( X_{\leq s} ). 
$$
Finally, we observe that the quasi-renewal property follows from the fact that the conditional distribution $X_{t} | X_{\leq t} - \mathbf{E}[X_t | X_{\leq s}]$ is zero-mean Gaussian with covariance that depends only on the distance to $t_0$, so both sides are equal in distribution.

\subsection{Proof of \Cref{thm:spectrum}}

\subsubsection{Instantenous Spectrum}

Since \Cref{eq:rlfbm} defines a nonstationary process, we are going to use the Wigner-Ville instantaneous spectrum, which is the time-varying Fourier transform of the covariance with respect to the lag:

\begin{equation}
\label{eq:wigner}
S_X(t,\omega)=\hat{F}_\tau[K_X(t-\tau/2,t+\tau/2)](\omega)=\int K_X(t-\tau/2,t+\tau/2)e^{-i\omega\tau}d\tau,
\end{equation}

We easily find that the covariance of $X=X_H$ in terms of time-difference coordinates equals

\begin{equation}
\label{eq:covariance}
K_X(t-\tau/2,t+\tau/2)=\frac{(t-|\tau|/2)^{H+1/2}(|\tau|/2+t)^{H-1/2}{_2F_1}(1,1/2-H;H+3/2;\frac{2t-|\tau|}{2 t+|\tau|})}{\Gamma(H+1/2)\Gamma(H+3/2)}
\end{equation}

when $|\tau|<t$, and is zero otherwise. Rather than directly transforming the covariance in \cref{eq:covariance}, we will be working with its derivative which happens to have a very simple form:

\begin{equation}
\label{eq:first_derivative}
\frac{\partial}{\partial t}K_X(t-\tau/2,t+\tau/2)=\frac{2^{1-2H}(4t^2-\tau^2)^{H-\frac{1}{2}}}{\Gamma\left(H+\frac{1}{2}\right)^2}\cdot\mathbf{1}_{\{|\tau|<2 t\}},
\end{equation}

and whose Fourier transform is easy to find in terms of the hypergeometric function $_0F_1$:

\begin{equation}
\label{eq:fourier_derivative}
\mathcal{F}_\tau[\frac{\partial}{\partial t}K_X(t-\tau/2,t+\tau/2)](\omega)=\frac{2\sqrt{\pi}\left(\frac{t}{\omega}\right)^HJ_H(2t\omega)}{\Gamma\left(H+\frac{1}{2}\right)}.
\end{equation}

By taking the antiderivative twice, we find that the spectrum can be expressed in terms of $_1F_2$:

\begin{equation}
\label{eq:spectrum}
S_X(t,\omega)=\frac{2^{2H+1}t^{2H+1}{_1F_2}(H+1/2;H+1,H+3/2;-t^2\omega^2)}{\Gamma(2H+2)}.
\end{equation}




The representation in \cref{eq:spectrum} has the advantage that the $_1F_2$ expression represents a smooth function which is easy to approximate. Using the theory of hypergeometric functions, we easily find that the instantaneous spectrum oscillates around $\omega^{-2H-1}$:

\begin{equation}
\label{eq:oscillation}
S_X(t,\omega)=\omega^{-2H-1}\left(1-O\left(\frac{(t\omega)^{H-1/2}(\sin(\pi H/2-2t\omega)+\cos(\pi H/2-2t\omega))}{2\sqrt{\Gamma(H+1/2)}}\right)\right).
\end{equation}

\subsubsection{Convergence of Time-Averaged Spectra}

Despite these growing oscillations, we can prove that the running average converges and matches the power-law. Namely, when $0<H<3/2$ we have the explicit formula

\begin{equation}
\label{eq:time_avg}
\frac{1}{T}\int_0^T S_X(t,\omega)\,dt=\frac{2^{2H+1}T^{2H+1}{_1F_2}(H+1/2;H+3/2,H+2;-T^2\omega^2)}{\Gamma(2H+3)},
\end{equation}

with the asymptotic behaviour

\begin{equation}
\label{eq:asymptotic}
\frac{1}{T}\int_0^T S_X(t,\omega)\,dt=\omega^{-2H-1}(1+O_H((T\omega)^{H-3/2})),\quad T\omega\gg 1,
\end{equation}

demonstrating, for $0<H<3/2$, the power law $\omega^{-2H-1}$ as expected.

It should be noted that using the asymptotic expansion \cref{eq:oscillation} is insufficient to obtain the convergence rate in \cref{eq:asymptotic}. This is because its second-order error term $O((t\omega)^{H-3/2})$ does not capture oscillations, as seen with $H=1/2$ resulting in an error term $O(\log T/(T\omega))$.

\end{document}